
\def\d#1/d#2{ {\partial #1\over\partial #2} }
\def\pdr{\partial}

\def\eps{\epsilon}
\def\half{{1\over 2}}
\def\tr{\hbox{tr}}

\def\BEGINIGNORE#1ENDIGNORE{}

\def\un#1{\underline{#1}}

\def\linebreak{\hfil\break}


\def\ifundefined#1{\expandafter\ifx\csname
#1\endcsname\relax}

\newcount\eqnumber \eqnumber=0
\def\beq{ \global\advance\eqnumber by 1 $$ }
\def\eeq{ \eqno(\the\eqnumber)$$ }
\def\label#1{\ifundefined{#1}
\expandafter\xdef\csname #1\endcsname{\the\eqnumber}
\else\message{label #1 already in use}\fi}
\def\(#1){(\csname #1\endcsname)}
\def\puteqno{\global\advance \eqnumber by 1 (\the\eqnumber)}

\newcount\refno \refno=0
\def\[#1]{\ifundefined{#1}\advance\refno by 1
\expandafter\edef\csname #1\endcsname{\the\refno}
\fi[\csname#1\endcsname]}
\def\refis#1{\noindent\csname #1\endcsname. }

\baselineskip=20pt
\parskip=10pt
\magnification=1200

UR-1266

ER40685-720

hep-th/9207039
\vskip.4in
\centerline { \bf Current Algebra in Three Dimensions}
\vskip.4in
\centerline{ G. Ferretti and S.G. Rajeev }
\vskip.4in
\centerline{\it Department of Physics and Astronomy}
\centerline{\it University of Rochester}
\centerline{\it Rochester, N.Y. 14627}
\vskip.9in
\centerline{ \bf Abstract}

\noindent
We study a three dimensional analogue of the Wess--Zumino--Witten model,
which describes the Goldstone bosons of three dimensional Quantum
Chromodynamics. The topologically non--trivial term of the action  can
also be viewed as a nonlinear realization of  Chern--Simons form.
We obtain the current algebra of this model by canonical methods.
This is a three dimensional generalization of the Kac--Moody algebra.
\vskip.5in


\vfill\eject

Three dimensional field theories with topological terms in the action have
been studied recently in many physical contexts  such as Chern--Simons
theories\[chsim], models for anti--ferromagnets\[fradkin] and anyon
statistics\[anyons]. (See \[reviews] for an overview). It is known that
topological terms affect the statistics of solitons \[anyons],\[witten],
and can change the canonical commutation relation of
observables\[rajeev1],\[mick],\[faddeev].
In three dimensions, this issue was studied
in refs. \[bowick], \[semenoff]. From
another direction, three dimensional nonlinear models have been shown to be
renormalizable\[largen]  in the $1/N$ expansion, although
they are not so by power counting.

In this paper we study a three dimensional nonlinear
sigma model on a coset space
with a topological term. It can be viewed as a three dimensional analogue of
the Wess--Zumino--Witten \[witten] model or as a  nonlinear Chern--Simons
theory\[chsim]. The model arises as the low energy limit of three dimensional
Quantum Chromodynamics (QCD) \[3dqcd]. It is also related to certain models for
anti--ferromagnetism that arise in attempts to explain high $T_c$
superconductivity\[fradkin]. There are solitons in this model whose statistics
is determined by the topological term. Furthermore,  the theory should be
renormalizable in the $1/N$ expansion. The focus of the present paper is the
canonical formalism of this model. We discover by this method  a three
dimensional generalization of the Kac--Moody algebra; it is a non--trivial
abelian extension of the naive current algebra. Like the Kac--Moody algebra,
this can be further extended to a semi--direct product with the algebra
of vector fields.

The Wess--Zumino--Witten (WZW) model \[witten] in an  even
dimensional space--time describes   anomalous  global symmetries. The field
variable $g$ takes values in a compact Lie group $G$ (typically $SU(N)$). It
satisfies the classical equation of motion (for two dimensions):
$$
       \pdr_\mu(g^{-1}\pdr^\mu g)+\lambda \eps^{\mu\nu}(g^{-1}\pdr_\mu g
       g^{-1}\pdr_\nu g)=0.
$$
If $\lambda=0$, the equation is invariant under two discrete symmetries,
$P_1:g\to g^{-1}$, and  $P_2:g(t,x)\to g(t,-x)$.
The WZW term breaks the symmetry down to the product $P_1P_2$.
Thus, the WZW term forces the fundamental field of
the theory to be a pseudo--scalar.
It is possible to formulate
\[witten],\[rajeev1],\[bhattraj], this theory in a
canonical formalism entirely in terms of currents. The classical Poisson
brackets of the currents then define an infinite dimensional Lie algebra, the
current algebra. In the absence of the WZW  term the current algebra is just
the set of maps from space to the Lie algebra of $G$,
with the point--wise bracket
\[rajeev1]. The current algebra is modified by the WZW term. In
general, it provides an extension of the current algebra by an abelian
algebra\[mick],\[faddeev]. In particular, in two dimensional
space--time, the current  algebra  is a central
extension of the loop algebra, the well-known affine Kac--Moody algebra. The
representation theory of this algebra is well--understood. The relation to WZW
model has clarified the representation theory by relating it to conformal field
theory.

Much less is known about the representation theory of current algebras in
higher dimensions. Some progress has been made in
this direction \[micraj],
although a complete understanding is still not available.
This motivated us to look for an
analogue of the WZW model in $2+1$
dimensions. This would be a way  to study current algebras and Bose--Fermi
equivalence in a context simpler than $3+1$ dimensions, yet more  general
than  $1+1$ dimensions.

However, there are no anomalies for continuous symmetries in odd
dimensional space--time. This is related to the fact that $H^{4}(G)$ vanishes
for the classical Lie groups. We can find an analogue by looking for a
nonlinear sigma model
on a target space with $H^{4}$ non--zero. Futhermore the additional term must
preserve parity if  the field variable is a pseudo--scalar.
The answer\[3dqcd] is the target space
$Gr_{n,N}=U(N)/U(n)\times U(N-n)$, the Grassmannian. We can parametrize
$Gr_{n,N}$ by an $N\times N$ matrix $\Phi$:
$$
        Gr_{n,N}=\{\Phi| \Phi^{\dag}=\Phi;\Phi^2=1;\tr \Phi=N-2n\}.
$$
The nonlinear model with this target space has the field
equation $[\Phi,\pdr_\mu\pdr^\mu \Phi]=0$.
The  cohomology group $H^4(Gr_{n,N})=Z\oplus Z$, (for $N\geq 4,n\geq 2 $) is
generated
by  $\omega_4$ and $\omega_2\wedge \omega_2$, where
$\omega_{2k}=\tr \Phi(d\Phi)^{2k}$.
Of the two generators, only $\omega_4$ is odd
under the transformation $\Phi\to -\Phi$.
Thus we arrive at a generalization of the WZW model to $2+1$ dimensions,
\beq
        F_{\pi}[\Phi,\pdr_\mu
\pdr^\mu\Phi]+{k\over{8\pi}}\eps^{\mu\nu\rho}
\pdr_{\mu}\Phi\pdr_{\nu}\Phi\pdr_\rho
\Phi=0.\label{eqmotion}
\eeq
The coupling constant $F_\pi$ has
dimension of inverse length in the classical theory. This equation of motion
follows from the multivalued action
\beq
S[\Phi]={{F_\pi}\over 2}
\int_M\tr d\Phi*d\Phi+{k\over 64\pi}\int_{M_4} \tr
\Phi(d\Phi)^4. \label{sphi}
\eeq
Here, $M_4$ is a four dimensional manifold whose boundary is space--time $M$.
As in the WZW model, in order  that $\exp(iS[\Phi])$ be independent of the
continuation into the fourth dimension, $k$ must be an integer. This theory is
invariant under parity (with  $\Phi$ transforming as a pseudo--scalar) if
$N=2n$. In this case it is the low energy limit of three dimensional QCD with
an even number $2n$ of flavors and $k$ colors \[3dqcd].
But we will study this theory for  general $N$ and $n$.

We will now present a canonical   formulation of this theory, in terms of a
set of Poisson brackets for the basic observables,  a set of first class
constraints and a Hamiltonian.
The Poisson brackets of the theory will be a generalization
of the Kac--Moody algebra to $2+1$ dimensions.
The canonical formalism is in terms of a decomposition of space--time
$M=\Sigma\times R$, $\Sigma$ being the two dimensional surface at fixed time.
It is possible to derive this  formalism from an action principle, but the
appropriate  one is not the  multi--valued action \(sphi). Define a new
variable $g$ valued in $G\equiv U(N)$.
One can always write $\Phi=g\eps g^{\dag}$
with $\eps={\rm diag}\{1,\cdots,1,-1,\cdots,-1\}$, and $\tr\eps=N-2n$.
Then $\Phi$ is invariant under right
multiplication of $g$ by elements that commute with $\eps$; i.e., under
$g\mapsto gh, h\in H\equiv U(n)\times U(N-n)$.
These transformations are therefore
like gauge transformations and we can write an action for the theory in terms
of $g$ that respects this gauge invariance:
\beq
  S[g,A]=-2F_\pi\int_M \tr \bigg((g^{\dag}dg-A)*
        (g^{\dag}dg-A)\bigg) -
        {k\over 8\pi}\int_M \tr \bigg(\eps(g^{\dag}dg)^3-{1\over
         3}(g^{\dag}dg\eps)^3\bigg). \label{Sga}
\eeq
The one form $A$ is an auxiliary gauge field valued in $\underline H$,
the Lie algebra of $H$. Its purpose is to remove the unwanted
degrees of freedom.
In this form of the  action, the topologically non--trivial term is a nonlinear
realization of the Chern--Simons term. Unlike \(sphi), action \(Sga)
is a local integral on space--time
but it is only gauge invariant up to a multiple of $2\pi$.

It is now possible to derive Poisson brackets
and constraints from this action by a conventional procedure.  The
Poisson brackets so obtained  at first will  involve  nonlinear (cubic) terms
in $\Phi$. However we can remove these by appropriate redefinition
of  the generators. The variables $\Phi,J$ that satisfy simple commutation
relations are, in this language, $\Phi=g\eps g^{\dag}$ and
$$
 J=g\bigg(F_\pi
 R+{k\over{16\pi}}\eps^{ij}(g^{\dag}\partial_i g
 g^{\dag}\partial_j g \eps+\eps g^{\dag}\partial_i g
 g^{\dag}\partial_j g -\eps g^{\dag}\partial_i g
 \eps g^{\dag}\partial_j g
 \eps)\bigg)g^{\dag}. \label{current}
$$
Here, $R$ is  the projection
of $g^{\dag}\dot g$ on the orthogonal complement of the gauge group and must
satisfy the constraint $[R,\eps]_+=0$. (The symbol $[\;\;\;]_+$ will denote
the anti-commutator throughout the paper). Our conventions are that $\Phi$ is
hermitian and $J$ anti--hermitian.

We  present only the results of the canonical analysis, leaving the details for
a longer publication.
The basic observables of the theory are $\Phi$ and $J$,
(which is essentially the time component of the current),
specified on $\Sigma$.
It is natural to think of $\Phi$ as a scalar on $\Sigma$ and to $J$ as a
scalar density (two--form). Let us also introduce the test functions
$\lambda:\Sigma\to \un{G}$ scalar, and $\xi:\Sigma\to \un{G}$ scalar density.
($\un{G}$ denotes the Lie algebra of $G$).
We give the Poisson brackets in terms of the dimensionless quantities
$\Phi(\xi)\equiv \int_\Sigma\tr(\Phi\xi) d^2x$
and $J(\lambda)\equiv\int_\Sigma\tr (J\lambda) d^2x$. They are
\beq\eqalign{
        &\{\Phi(\xi),\Phi(\xi')\}=0\quad \quad\hbox{,}\quad
        \{J(\lambda),\Phi(\xi)\}=\Phi([\lambda,\xi]),
        \quad\hbox{and}\cr
        &\{J(\lambda),J(\lambda')\}=J([\lambda,\lambda'])+
        k\Phi(\omega(\lambda,\lambda')).\cr\label{lie}}
\eeq
In \(lie), $\omega$ is defined as $\omega(\lambda,\lambda')=(1/16\pi)
\eps^{ij}[\pdr_i\lambda,\pdr_j\lambda']_+$.
If the space $\Sigma$ is a torus,
we can  write these relations more explicitly in a
plane wave basis:
\beq\eqalign{
   &\{\Phi^a_m,\Phi^b_n\}=0\quad\hbox{,}\quad
   \{J^a_m,\Phi^b_n\}=f^{abc}\Phi^c_{m+n},\quad\hbox{and} \cr
   &\{J^a_m,J^b_n\}=f^{abc}J^c_{m+n}-
   {k\over 16\pi} d^{abc}\eps^{ij}m_in_j\Phi^c_{m+n}.\label{liecomp}\cr}
\eeq
In \(liecomp), $m,n$ are two-dimensional vectors with integer components.
Also, $d^{abc}$ is the usual symmetric cubic invariant of $U(N)$ and
$f^{abc}$ the structure constants.

The algebra has to be supplemented by two constraints
\beq
        \Phi^2=1\quad\hbox{and}\quad
         [J,\Phi]_+ +{k\over 16\pi}\epsilon^{ij}
  (\partial_i\Phi\partial_j\Phi)=0. \label{constraint}
\eeq
One can verify that these are first class constraints.
This is one major difference between our treatment of the problem and
the usual canonical formalism for similar models \[bowick].
With our method, second class constraints never arise and there is no need
to introduce Dirac brackets. Actually, our
constraints \(constraint) satisfy even stronger relations that the
conditions for being first class. It can be easily checked from \(lie)
that the Poisson brackets of $\Phi$ and $J$ with \(constraint) vanish weakly.
This means that every two weakly equivalent observables $A\approx B$
of the theory will have weakly equivalent Poisson brackets with any
third observable $C$: $\{A,C\}\approx \{B,C\}$.
Both constraints and Poisson brackets are also invariant
under diffeomorphisms of $\Sigma$.

The canonical formalism is completed by the Hamiltonian
\beq
 H=\half\int_{\Sigma}\tr\bigg(-{1\over{F_\pi\sqrt{g}}}
 \big(J+{k\over 32\pi}\epsilon^{ij}\partial_i\Phi
 \partial_j\Phi\big)^2+
 {{F_\pi\sqrt{g}}\over 4}g^{ij}\partial_i\Phi\partial_j\Phi\bigg)d^2x.
\eeq
Of course, the hamiltonian does depend on the metric $g_{ij}$
on $\Sigma$. The equations of motion that follow are
\beq\eqalign{
        \dot \Phi&={1\over{F_\pi\sqrt{g}}}[J,\Phi]\cr
        \dot J&=- {1\over 4}F_\pi\sqrt{g}g^{ij}[\Phi,
        \pdr_i\pdr_j\Phi]+{k\over {32\pi F_\pi\sqrt{g}}}
        \eps^{ij}
        \bigg([\pdr_iJ,\pdr_j\Phi]_+-\pdr_i(\Phi J\Phi)\pdr_j\Phi-
        \pdr_j\Phi\pdr_i(\Phi J\Phi)\bigg).\cr
}\eeq
The equation \(eqmotion) for $\Phi$ then follows (in flat space) from
this system of first order equations and the constraints.
This completes our discussion of the canonical formalism.

Equations \(lie), (or \(liecomp)), define
our current algebra ${\cal G}_k$ (more exactly
a current--field algebra). Notice that
the Poisson brackets yield {\it linear} relations in $\Phi$ and $J$,
so \(lie) defines a Lie algebra.
If we set $k=0$, the $J$'s alone form a subalgebra ${\cal J}$. The
$\Phi$'s generate an abelian sub--algebra ${\cal V}$ of $\un{G}$-valued
densities. The vector space ${\cal V}$
can be identified with dual of the Lie algebra
${\cal J}$ by the natural pairing
$<\lambda,\xi>=\int_\Sigma \tr (\lambda\xi) d^2x$,
so that it carries the co--adjoint representation of ${\cal J}$. When
$k=0$, our algebra ${\cal G}_0$
reduces to the semi--direct product of ${\cal J}$ with its
co--adjoint representation.

When $k\neq 0$, ${\cal G}_k$ is an abelian extension of the map algebra
${\cal J}$ by its co--adjoint representation. The Jacobi identity of
${\cal G}_k$ is equivalent to the statement that $\omega:{\cal J}\wedge{\cal
J}\to {\cal V}$ is a two--cocycle of the Lie algebra cohomology:
\beq
    \pdr\omega(\lambda_1,\lambda_2,\lambda_3)\equiv
    [\lambda_1,\omega(\lambda_2,\lambda_3)]+
   \omega(\lambda_1,[\lambda_2,\lambda_3])+{\rm cyclic}=0,
\eeq
which can be verified by direct computation. If $\omega$ had been exact there
would have been a linear function $\mu:{\cal J}\to {\cal V}$ such that
\beq
\omega(\lambda,\lambda')=\pdr\mu(\lambda,\lambda')
\equiv -\mu([\lambda,\lambda'])+
[\lambda,\mu(\lambda')]-[\lambda',\mu(\lambda)].
\eeq
There is no such $\mu$, so that we cannot reduce our algebra to a semi--direct
product by a change of basis; ${\cal G}_k$ is a non--trivial abelian
extension of ${\cal J}$ by ${\cal V}$.

It is useful to note that the above extension can be `exponentiated'
to an extension of the group of maps $g :\Sigma \to G$ by the vector space
${\cal V}$, (thought of as an abelian group). The multiplication law $\circ$ is
\beq
        (g_1,\xi_1)\circ (g_2,\xi_2)=(g_1g_2,\xi_1+g_1\xi_2 g_1^{-1}
        +k\Omega(g_1,g_2)),
\eeq
where $\Omega(g_1,g_2)=(1/16\pi) \eps^{ij}\pdr_ig_1\pdr_j
g_2 g_2^{-1}g_1^{-1}$.
The associativity of the group multiplication $\circ$ requires that
$\Omega$ be a group 2-co--cycle:\[mickbook]
\beq
    \pdr\Omega(g_1,g_2,g_3)\equiv -\Omega(g_1g_2,g_3)+\Omega(g_1,g_2g_3)-
    \Omega(g_1,g_2)+g_1\Omega(g_2,g_3)g_1^{-1}=0.
\eeq
This identity can be proved by direct computation.
It is possible to understand the constraints
\(constraint) as describing a co--adjoint orbit
of the above group. The symplectic form and hence the part of the action that
is linear in time derivatives can be understood from Kirillov's method of
orbits applied to this case. In fact, it turns out that the
Poisson bracket structure derived in this way coincides with \(lie).

Finally, recall that the Kac--Moody algebra is invariant
under the action of the Virasoro algebra. In fact the generators
of the Virasoro algebra can be written in terms of the currents.
The analogue of the Virasoro algebra in our case is the algebra of vector
fields on $\Sigma$.
The Lie algebra ${\cal G}_k$ is invariant under diffeomorphisms, so
that it can be extended as a semi--direct product with the algebra of vector
fields on $\Sigma$. If $u$ and $v$ are such vector fields, and $L$ is the
generator associated to them, satisfying $\{L(u),L(v)\}=L([u,v])$, then,
\beq
        \{L(u),J(\lambda)\}=J(u^i\pdr_i\lambda)\quad\hbox{and}
        \quad \{L(u),\Phi(\xi)\}=\Phi(\pdr_i(u^i \xi)).
\eeq

It would be interesting to develop a representation theory for the algebra
\(liecomp).
Physically, that would correspond to quantizing the above  field theory.
This might look impossible at first because the theory is not renormalizable
by power counting. However,
as remarked in \[3dqcd], the theory is renormalizable in the $1/N$
expansion, provided one allows for massive vector fields to be dynamically
generated. We would first consider the limit
$N\to \infty$ (keeping $n$ and $k$
fixed)  that is  solvable by  the saddle point method. This model
has a non--trivial UV fixed point.
This is the analogue of the UV fixed point of
the WZW model in 1+1 dimensions (although in that case the UV
fixed point is trivial).
The WZW model also has a non--trivial
IR stable fixed point. It is possible that there is
an analogous ({\it non--trivial}) IR fixed point in our theory as well.

Another interesting issue is that of Fermi--Bose correspondence. In two
dimensions the WZW model,  at the IR fixed point, corresponds to a free Fermi
theory. There has already been an attempt to prove such an equivalence for the
$CP^{1}$ model \[kovner]. However there is,
at present, no reliable approximation
method to study this issue, because the correspondence breaks down for
$CP^N$ with $N>1$.
Our model should have Fermionic equivalents for any $N,n$, so that the
issue can be studied within the $1/N$ expansion. We will report on work
in this direction in a later publication.

We would like to thank A.P. Balachandran, P. Teotonio and Z. Yang for
discussions. This work was supported in part by the US Department of Energy,
Contract No. DE-AC02-76ER13065.

\vfill\eject

\centerline{\bf References}

\refis{chsim}
S. Deser, R. Jackiw and S. Templeton. Phys. Rev. Lett. {\bf 48}, 975 (1982);
\hfill\break
S. Deser, R. Jackiw and S. Templeton. Ann. Phys. {\bf 140}, 372 (1982);
\hfill\break
J. Schonfeld, Nucl. Phys. {\bf B185}, 157 (1981);
\hfill\break
E. Witten, Commun. Math. Phys. {\bf 121}, 351 (1989).

\refis{fradkin} E. Fradkin, ({\it Field Theories of Condensed Matter Systems}
Addison--Wesley, New York, 1991);
\hfill\break
P.B. Wiegmann, Phys. Scripta, {\bf T27}, 160 (1989);
\hfill\break
N. Read and S. Sachdev, Nucl. Phys. {\bf B316}, 609 (1989).

\refis{anyons} F. Wilczek and A. Zee, Phys. Rev. Lett. {\bf 51}, 2250 (1983).

\refis{reviews} S. Forte, Rev. Mod. Phys. {\bf 64}, 193 (1992);
\hfill\break
A.P. Balachandran, M. Bourdeau and S. Jo, Int. J. Mod. Phys.
{\bf A5}, 2423 (1990).

\refis{witten} E. Witten, Nucl. Phys. {\bf B223}, 422 (1983);
Comm. Math. Phys. {\bf 92}, 455 (1984).

\refis{rajeev1} S.G. Rajeev, Phys. Rev. {\bf D29}, 2944 (1984).

\refis{mick} J. Mickelsson, Comm. Math. Phys. {\bf 97}, 361 (1985).

\refis{faddeev} L.D. Faddeev, Phys. Lett. {\bf B145}, 81 (1984).

\refis{bowick} M.J. Bowick, D. Karabali and L.C.R.
Wijewardhana, Nucl. Phys. B271 417 (1986).

\refis{semenoff} G. Semenoff, P. Sodano and Y.S. Wu, Phys. Rev. Lett.
{\bf 62}, 715 (1989);
\hfill\break
G. Semenoff, Phys. Rev. Lett. {\bf 61}, 517 (1988);
\hfill\break
R. Banerjee, Phys. Rev. Lett. {\bf 69}, 17 (1992).

\refis{largen}
I.Y. Aref'eva and S.I. Azakov,
Nucl. Phys. {\bf B162}, 298 (1980);
\hfill\break
B. Rosenstein, B.J. Warr and S.H. Park, Nucl. Phys. {\bf B336}, 435 (1990);
\hfill\break
A. Kovner and B. Rosenstein, Phys. Lett. {\bf B261}, 104 (1990).

\refis{3dqcd} G. Ferretti, S.G. Rajeev and Z. Yang, {\it The Effective
Lagrangian of Three Dimensional Quantum Chromodynamics}, U. of
Rochester Preprint
UR 1255 (1992) to appear in Int. J. Mod. Phys.;
\hfill\break
{\it  Baryons as Solitons in
Three Dimensional Quantum Chromodynamics}, U. of Rochester
Preprint UR 1256 to appear in Int. J. Mod. Phys.

\refis{bhattraj} G. Bhattacharya and S.G. Rajeev, Nucl. Phys. {\bf B246},
157 (1984).

\refis{micraj} J. Mickelsson and S.G. Rajeev, Comm. Math. Phys. {\bf 116},
365 (1988); Lett. Math. Phys. {\bf 21}, 173 (1991).

\refis{mickbook} J. Mickelsson, ({\it Current Algebras and Groups},
Plenum, New York, 1989).

\refis{kovner} A. Kovner, Int. J. Mod. Phys., {\bf A5}, 3999 (1990).
\bye